\documentclass[reprint,prl,showpacs,aps,amsmath,amssymb,nofootinbib,floatfix,floats,superscriptaddress]{revtex4-1} 

\allowdisplaybreaks[2]
  
\usepackage{graphicx}
\usepackage[usenames, dvipsnames]{color}
\usepackage[colorlinks, pdfborder={0 0 0}, plainpages=false]{hyperref}
\usepackage{breakurl}
\usepackage{amsmath, amssymb, amsfonts}
\usepackage{xspace} 
\usepackage{dcolumn}
\usepackage{bm}
\usepackage{multirow} 
\usepackage{graphics,psfrag}
\usepackage{graphicx,psfrag}

\definecolor{CiteColor}{rgb}{0, 0.5, 0}
\hypersetup{citecolor=CiteColor} %
\definecolor{RefColor}{rgb}{0.55, 0, 0}
\hypersetup{linkcolor=RefColor} %

\usepackage{ulem}

\usepackage{amsmath}
\usepackage{times}
\usepackage{mathptmx}

\definecolor {darkgreen}{rgb}{0.2, 0.7, 0.2}

\newcommand{\Maryland}{\affiliation{Maryland Center for Fundamental
    Physics \& Joint Space-Science Institute,\\ Department of Physics,
    University of Maryland, College Park, MD 20742, USA}}
\newcommand{\Caltech}{\affiliation{Theoretical Astrophysics 350-17,
    California Institute of Technology, Pasadena, CA 91125, USA}}
\newcommand{\Cornell}{\affiliation{Center for Radiophysics and Space
    Research, Cornell University, Ithaca, New York 14853, USA}}
\newcommand{\CITA}{\affiliation{Canadian Institute for Theoretical
    Astrophysics, 60 St.~George Street, University of Toronto,
    Toronto, ON M5S 3H8, Canada}} %
\newcommand{\CIFAR}{\affiliation{Canadian Institute for Advanced Research, 180 Dundas St.~West, Toronto, ON M5G 1Z8, Canada}} %

\newcommand{\Adj}[2]{A^{(#1)}_{#2}}
\newcommand{\Acal}[2]{\bar{A}^{(#1)}_{#2}}

\begin{document}

\title{Stability of nonspinning effective-one-body model in approximating two-body dynamics and gravitational-wave emission}

\author{Yi Pan} \Maryland %
\author{Alessandra Buonanno} \Maryland %
\author{Andrea Taracchini} \Maryland %
\author{Michael Boyle} \Cornell 
\author{Lawrence E. Kidder} \Cornell 
\author{Abdul H.~Mrou\'{e}} \CITA 
\author{Harald P.~Pfeiffer} \CITA \CIFAR 
\author{Mark A.~Scheel} \Caltech 
\author{B\'{e}la Szil\'{a}gyi} \Caltech 
\author{Anil Zenginoglu} \Caltech

\begin{abstract}
The detection of gravitational waves and the extraction of physical information from them requires 
the prediction of accurate waveforms to be used in template banks. For that purpose, 
the accuracy of effective-one-body (EOB) waveforms has been improved over the last years by calibrating 
them to numerical-relativity (NR) waveforms. So far, the calibration has employed 
a handful of NR waveforms with a total length of $\sim 30$ cycles, the length being limited 
by the computational cost of NR simulations. Here we address the outstanding 
problem of the stability of the EOB calibration 
with respect to the length of NR waveforms. Performing  
calibration studies against NR waveforms of nonspinning black-hole binaries with 
mass ratios $1$, $1.5$, $5$, and $8$, and with a total length of $\sim 60$ 
cycles, we find that EOB  waveforms calibrated against either $30$ or 
$60$ cycles will be indistinguishable by the advanced detectors LIGO and Virgo when 
the signal-to-noise ratio (SNR) is below $110$. When extrapolating to a very large number of 
cycles, using very conservative assumptions, we can conclude that state-of-the-art 
nonspinning EOB waveforms of {\it any} length are sufficiently accurate for parameter estimation with advanced detectors 
when the SNR is below $20$, the mass ratio is below 5 and total mass is above $20 M_\odot$. 
The results are not conclusive for the entire parameter 
space because of current NR errors.
\end{abstract}

\date{\today}

\pacs{04.25.D-, 04.25.dg, 04.25.Nx, 04.30.-w}

\maketitle

{\it Introduction.} Coalescing compact-object binaries are among the most
promising gravitational-wave (GW) sources for ground-based
interferometric detectors such as LIGO, Virgo and KAGRA~\cite{Shoemaker2009,AdV,Somiya:2011np}. 
Accurate waveform models are crucial for detecting the signals and measuring the
physical parameters of the sources. By solving the Einstein equations 
numerically~\cite{Centrella:2010}, it is possible to produce 
accurate waveforms for the very late inspiral, merger and ringdown 
stages of the coalescence process. However, the length of numerical-relativity (NR) 
simulations is limited by their high computational cost, and today it is unrealistic to generate sufficiently many NR
waveforms long enough to be used directly in GW searches. 
The post-Newtonian (PN) formalism~\cite{Blanchet2006} is a slow-motion,
weak field approximation to the Einstein field equations that provides 
reliable  low-frequency inspiral waveforms. However, the PN approach becomes  
increasingly inaccurate close to merger~\cite{Buonanno:2009}. 
Several studies~\cite{Damour:2010,*Boyle:2011dy, *OhmeEtAl:2011, *MacDonald:2011ne, 
*MacDonald:2012mp} showed that there is a substantial gap between 
the frequency $f_{\rm PN}$ where PN waveforms cease being accurate and the frequency
$f_{\rm NR}$ where NR simulations start being available. The width of
the frequency gap $f_{\rm NR}\mbox{--}f_{\rm PN}$ depends on source parameters
and it is generally believed to increase rapidly with increasing mass ratio and
spin magnitudes. Much longer NR simulations can reduce $f_{\rm NR}$
while knowledge of higher-order PN terms in the two-body dynamics and 
radiation-reaction force can increase $f_{\rm PN}$~\cite{MacDonald:2012mp}, 
but it is extremely challenging to achieve those goals. 
An accurate description of the waveform in the frequency gap is thus 
an outstanding and pressing problem of GW source modeling, especially 
because advanced detectors will be operational in a few years.

The effective-one-body formalism~\cite{buonanno99,*buonanno00} (EOB) is a
successful approach that provides a complete description of the
coalescence of compact-object binaries. It uses the PN-expanded results in a
resummed form and incorporates results of black-hole perturbation
theory to produce waveforms for the inspiral, merger and ringdown stages. 
By construction, the EOB model reduces to the PN approximation at low 
frequency, while in the strong-field regime it models the merger and ringdown 
signals using physically motivated guesses and insights from perturbation theory. 
Following the breakthrough in merger simulations in NR~\cite{Pretorius2005a,*Campanelli-Lousto-Zlochower:2006,*Baker2006a}, the EOB model has been improved 
by calibrating it to progressively more accurate and longer NR simulations, 
spanning also larger regions of the parameter space~\cite{PanEtAl:2011,Taracchini2012,
Damour:2012ky,Pan:2013rra,Hinder:2013oqa}. Considering the success in calibrating NR waveforms, we expect that the EOB model will be able to interpolate/extrapolate 
NR waveforms over the entire source parameter space.
However, it is not yet clear whether the EOB calibration is stable under 
variation of the length of the NR waveforms that are used to calibrate the 
model, and whether EOB waveforms of length larger than the one used for 
calibration can safely be used 
to detect GW signals and extract physical parameters with advanced detectors. 

In this paper, we focus on the low-frequency, inspiral
  performance of the EOB model and assume, based on previous
  calibrations, that calibrated EOB merger and ringdown waveforms 
can be made indistinguishable from the NR ones~\cite{Littenberg:2012uj}. 
The EOB adjustable parameters that are used to calibrate the model 
not only improve EOB waveforms at high frequency, so that they match 
NR waveforms very accurately above $f_{\rm NR}$, but they also introduce 
deviations from known PN results in the frequency gap $f_{\rm PN} \mbox{--} 
f_{\rm NR}$. Below $f_{\rm PN}$ all PN-waveform families and the EOB waveforms 
agree with each other. The goal of this paper is to understand the accuracy 
of the EOB waveforms in the frequency gap, addressing the following questions: 
Is the EOB calibration stable with respect to the length of NR waveforms (i.e.,  
with respect to varying $f_{\rm NR}$)? If the calibration is stable when using  
the current length of NR simulations, for which we still have $f_{\rm
    NR} \gg f_{\rm PN}$, can we conclude that the calibrated EOB waveforms will be 
indistinguishable from the exact ones for all frequencies below $f_{\rm NR}$?

{\it Calibrating the effective-one-body model.} We calibrate the
EOB model against four nonspinning binary black-hole waveforms  
with mass ratios $q=1$, $1.5$, $5$ and $8$.  The $q=1$ simulation
was first presented in~\cite{MacDonald:2012mp}, and all four simulations are
presented in~\cite{Mroue:2013xna}. Table~\ref{tab:NR} lists 
the total number of GW cycles of the NR waveforms up to merger and including 
the junk radiation, and the maximum number of cycles $N_{\rm max}$ that we use when 
calibrating the EOB model (i.e., after removing the junk radiation). 
We decompose the EOB waveforms in -2 spin-weighted spherical-harmonic modes $(\ell,m)$. 
Previous studies~\cite{PanEtAl:2011} have shown that during the inspiral stage the frequency of all modes is well 
approximated by the $m$ multiple of the orbital frequency. Therefore, here 
for simplicity, we consider only the dominant $(\ell=2,m=2)$ mode. We expect that 
the results of our study hold also for the other modes since phase evolution of 
every mode is synchronized with the orbital phase.

\begin{table}
    \begin{tabular}{|c|cccc|}
      \hline
      $\quad q \quad$ & $\quad 1 \quad$ & $\quad 1.5 \quad$ & $\quad 5 \quad$ & $\quad 8 \quad$ \\
      \hline
      $N_{\rm sim}$ & $65$ & $66$ & $58$ & $52$ \\ 
      $N_{\rm max}$ & $60$ & $60$ & $55$ & $50$ \\
      \hline
    \end{tabular}
  \caption{\label{tab:NR} Total number of GW cycles $N_{\rm sim}$ 
of NR simulations (including junk radiation) up to merger 
and maximum number of cycles $N_{\rm max}$ used for EOB-model calibration, 
i.e., without junk radiation.}
\end{table}

The EOB inspiral-plunge dynamics for quasi-circular orbits is described by a set of 
Hamilton equations that include a dissipative force proportional to the rate of loss of 
the orbital energy. One then introduces adjustable parameters, i.e., unknown, 
higher-order PN terms, to improve both conservative 
and dissipative parts of the dynamics. To match EOB to NR waveforms within the NR error, only a few adjustable parameters are needed and their choice 
is not unique. In the nonspinning limit, the EOB  model depends only on two (or even one~\cite{Taracchini2013}) 
adjustable parameters $\Adj{i}{}$, $i=1,2$. We follow the parametrization of Ref.~\cite{Taracchini2012}, 
where two adjustable parameters were used in the nonspinning sector. The EOB inspiral waveform of mass ratio $q$ is therefore 
determined by the pair $\{\Adj{1}{},\Adj{2}{}\}$, where these coefficients depend on the mass ratio $q$.  
We calibrate the EOB model by mapping the phase difference between EOB
and NR waveforms in the $\Adj{1}{}$--$\Adj{2}{}$ parameter space, taking
into account NR errors in the simulations.

In our calibration procedure, we measure the phase difference at the end of 
inspiral, after aligning the EOB and NR waveforms at low frequency by shifting the EOB waveform
in time and phase. We determine the time and phase shifts
$\bar{t}_0$ and $\bar{\phi}_0$ by minimizing the square of the
difference between the GW phases of the NR and EOB waveforms
\begin{equation}\label{align}
\int_{t_1}^{t_2}\left[\phi_{22}^{\rm EOB}(t+t_0)+\phi_0-\phi_{22}^{\rm NR}(t)\right]^2\,dt \,,
\end{equation}
with respect to $t_0$ and $\phi_0$. The phase difference at a given time is given by
\begin{equation}\label{Dphit}
\Delta\phi(t) = \phi_{22}^{\rm EOB}(t+\bar{t}_0)+\bar{\phi}_0-\phi_{22}^{\rm NR}(t)\,,
\end{equation}
where $\bar{t}_0$ and $\bar{\phi}_0$ are the alignment parameters that minimize Eq.~\eqref{align}. 
The global phase difference over a time window $(t_1,t_3)$ is defined as
\begin{equation}
\Delta\phi_g = \max_{t\in(t_1,t_3)}|\Delta\phi(t)|\,.
\end{equation}
We set $t_3$ to the time of merger, i.e., to the time at which
$|h_{22}^{\rm EOB}|$ reaches its maximum. Because of NR errors
in $\phi_{22}^{\rm NR}$, the time shift $t_0$ and the global phase
difference $\Delta\phi_g$ are rather sensitive to the choice of the
time window $(t_1,t_2)$. To alleviate the effect of NR errors,
we choose $(t_1,t_2)$ following the prescription of Ref.~\cite{Taracchini2012}. 
We also repeat the alignment using four different choices of $(t_1,t_2)$ to estimate the uncertainty of $\Delta\phi_g$ due to NR errors.
\begin{figure*}
  \includegraphics[width=0.31\textwidth]{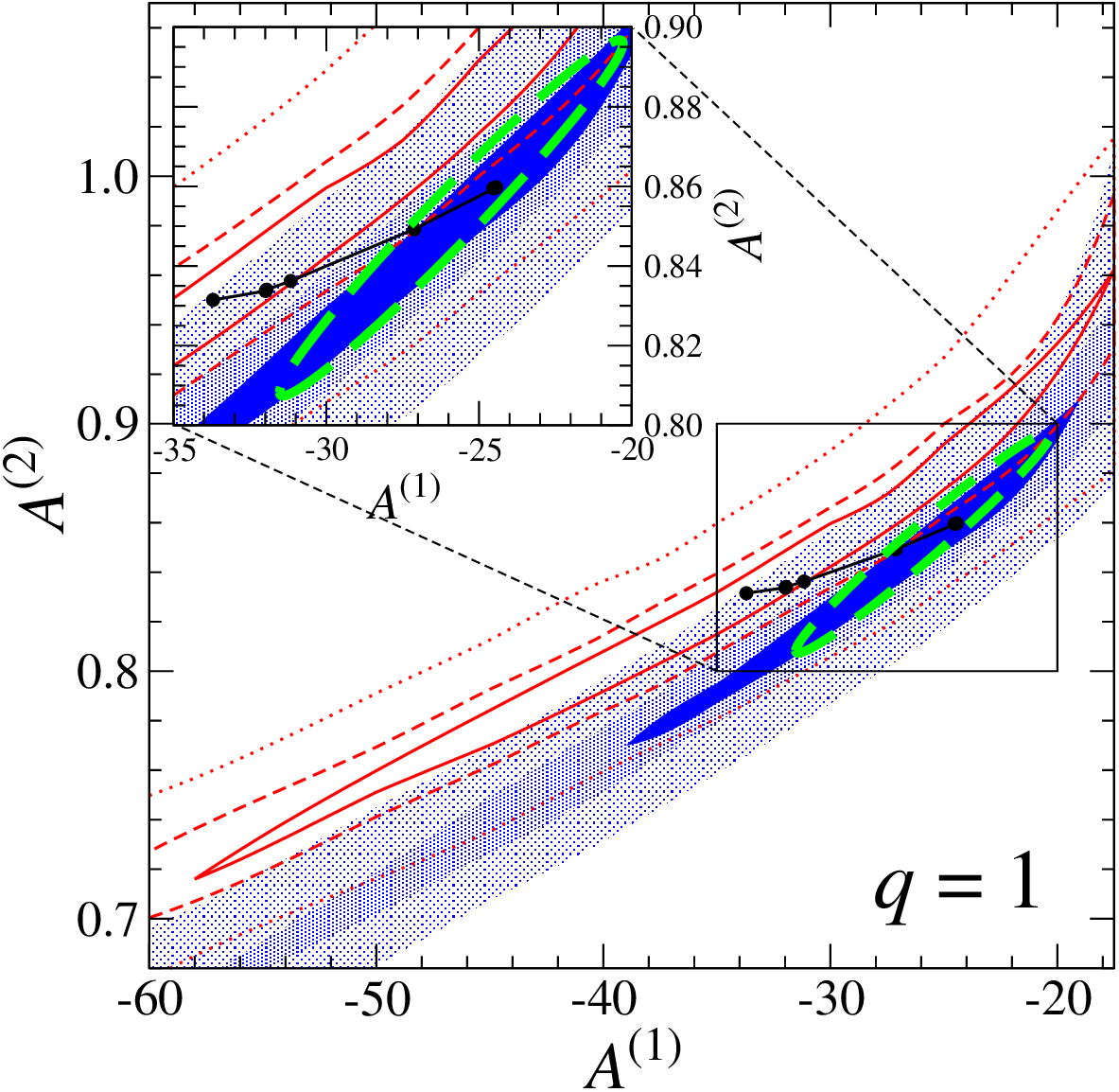}\quad
  \includegraphics[width=0.31\textwidth]{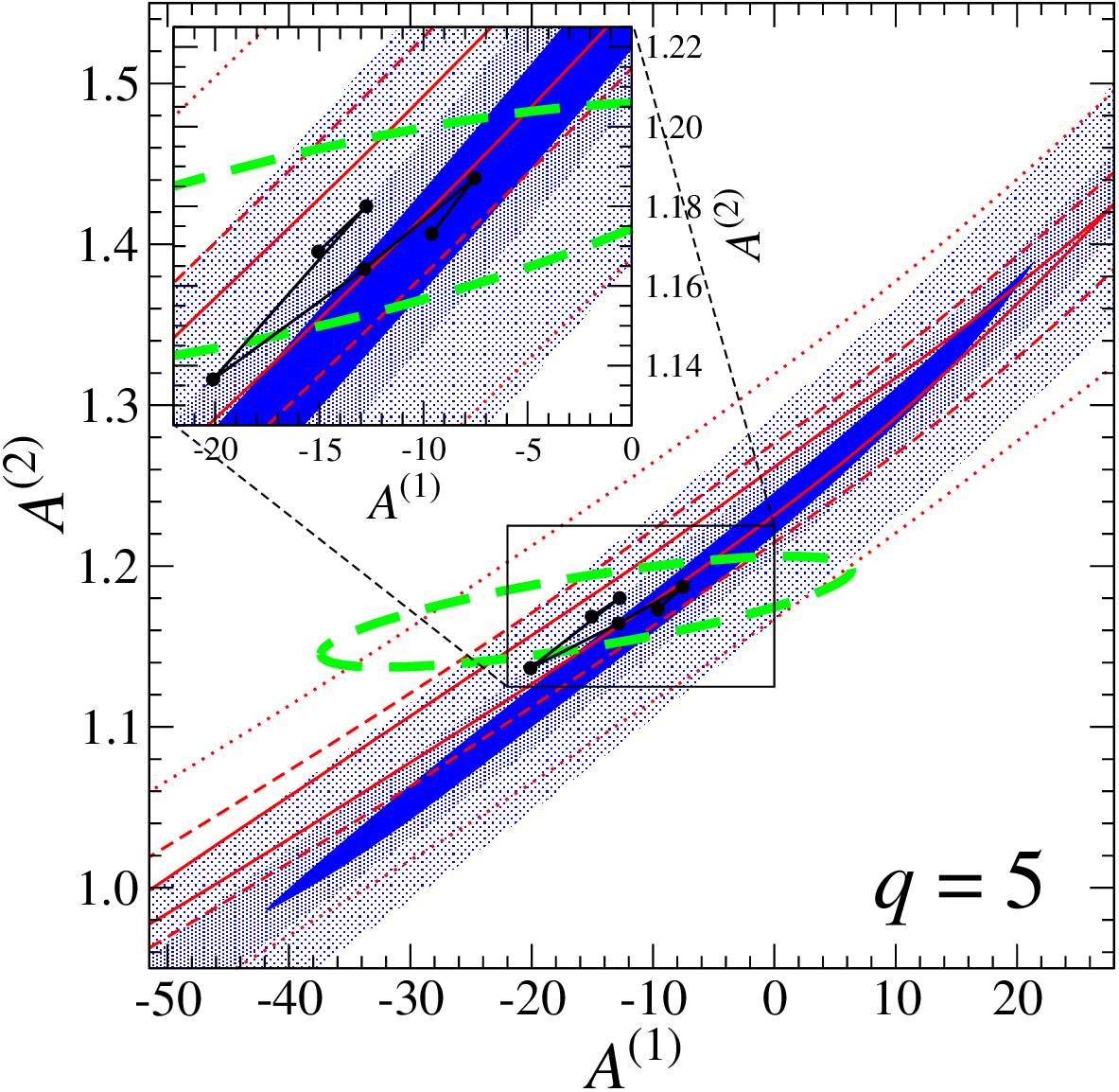}\quad
  \includegraphics[width=0.31\textwidth]{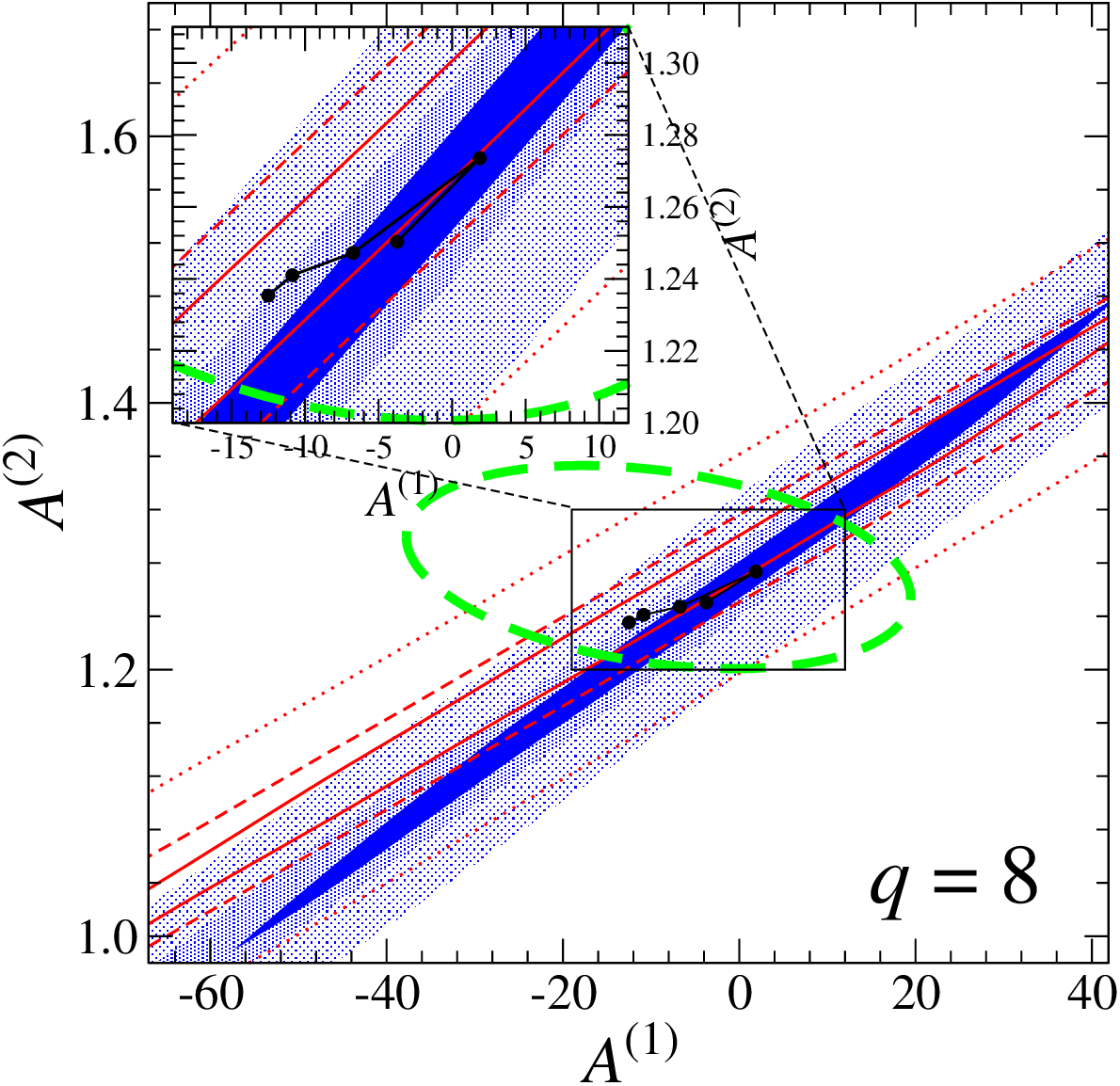}
  \caption{ \label{fig:contour} 
Contours of global phase
    difference $\Delta\phi_g$ between nonspinning NR waveforms of $N$
    GW cycles and EOB waveforms with adjustable parameters
    $\{\Adj{1}{},\Adj{2}{}\}$. The three panels show results for mass ratios
    $q=1$, $5$ and $8$ (from left to right). The shaded regions, from inside out, are
    $0.1$, $0.2$ and $0.5$ radian contours for comparisons with
    $N_{\rm max}$ cycles of NR waveforms. The solid, dashed and dotted
    lines are the same contours for comparisons with $30$ cycles of NR
    waveforms. The connected black dots are the calibrated points
    $\{\Acal{1}{N}(q),\Acal{2}{N}(q)\}$ for $N$ values changing from
    $30$ to $N_{\rm max}$.  
The inset zooms around these points.
    The NR error box of the calibrated point
    $\{\Acal{1}{N_{\rm max}}(q),\Acal{2}{N_{\rm max}}(q)\}$ is 
    show with the dashed ellipse.
  }
\end{figure*}
To calibrate the EOB model, we find those parameters
$\{\Acal{1}{N}(q),\Acal{2}{N}(q)\}$ that minimize $\Delta\phi_g$.  The
subscript $N$ indicates that calibration was performed using the last
$N$ GW cycles, i.e. $t_1$ corresponds to a time $N$ cycles before
merger, and $t_3$ is at merger. When building a calibrated
EOB model~\cite{Taracchini2012}, we fit the calibrated points
$\{\Acal{1}{N}(q),\Acal{2}{N}(q)\}$ to a smooth function in
$q$. However, since the fits' residuals are typically smaller than the
NR errors, we use here the calibrated points instead of the fitted
functions. We then increase $N$ from $30$ to $N_{\rm max}$ with a step
size of $5$ and determine how the point
$\{\Acal{1}{N}(q),\Acal{2}{N}(q)\}$ moves in the parameter
space. Besides systematics errors in the EOB model, the calibration
point can change also because of the NR errors.

The NR errors affect $\{\Acal{1}{N}(q),\Acal{2}{N}(q)\}$ in two
ways. The oscillatory phase errors at low frequency (due to residual eccentricity) 
introduce uncertainties in the alignment procedure, while the secular phase errors 
introduce uncertainties directly in the global phase difference $\Delta\phi_g$. To estimate
the impact of those NR errors on $\{\Acal{1}{N}(q),\Acal{2}{N}(q)\}$, we 
calculate those calibrated points using four different choices of the alignment time
window $(t_1,t_2)$ and three numerical waveforms: (i) the high
resolution, extrapolated to infinity with polynomial degree 3, 
(ii) the high resolution, extrapolated to infinity with polynomial degree 4, 
and (iii) the medium resolution, extrapolated to infinity with polynomial degree 3.
The differences between these numerical
waveforms represent the typical truncation and extrapolation
errors. Since we are only interested in the position (mean) and spread (variance) of $\{\Acal{1}{N}(q),\Acal{2}{N}(q)\}$, we do not investigate 
higher central moments and assume for simplicity a 
bivariate normal distribution of $\{\Acal{1}{N}(q),\Acal{2}{N}(q)\}$. 
We use the 12 data points to calculate the maximum likelihood estimators 
of their mean and variance. 

We summarize our results in Fig.~\ref{fig:contour} for $q = 1, 5,8$ and omit the
  $q=1.5$ case because it is very similar to the $q=1$ case. 
  When $N$ increases from $30$ to $N_{\rm max}$, the
  volume enclosed by the $\Delta\phi_g$ contours decreases gradually,
  reflecting tighter constraints from the calibration against longer
  NR simulations. Somewhat unexpectedly, the contours also shift and
  rotate smoothly, indicating a possible systematic change of the
  calibrated EOB model. For clarity, we show in Fig.~\ref{fig:contour} only the contours of $N=30$
  and $N_{\rm max}$ calibrations. In the inset of each panel, we zoom
  in around the calibrated points $\{\Acal{1}{N}(q),\Acal{2}{N}(q)\}$ to
  show their path when $N$ changes from $30$ to $N_{\rm max}$. We show
  also the NR error box of $\{\Acal{1}{N_{\rm
      max}}(q),\Acal{2}{N_{\rm max}}(q)\}$, which is the symmetric
  $95\%$ quantile of the estimated bivariate normal distributions. 
  In the $q=1$ case, the systematic drift of 
  $\{\Acal{1}{N}(1),\Acal{2}{N}(1)\}$ with increasing
  $N$ is not fully accounted for by the NR errors. Of course, it
  is in principle possible to improve the accuracy of the EOB model by
  calibrating it to the $N_{\rm max}$-cycle numerical
  waveforms. However, since the systematic differences between
  $\{\Acal{1}{N}(1),\Acal{2}{N}(1)\}$ are not much larger than the NR
  error boxes, the NR
  waveforms have to be as accurate as the $q=1$ waveforms employed in this paper
  to bring new information to the EOB calibration. For instance, the
  calibrated point $\{\Acal{1}{30}(1),\Acal{2}{30}(1)\}$ sits on the
  $0.5$-radian contour of $\Delta\phi_g$ obtained from the $N=60$
  calibration. That is to say, aligning a $q=1$, $60$-cycle NR
  waveform with a $60$-cycle EOB waveform generated by a model
  calibrated to a $30$-cycle NR waveform, such as the EOB model in 
  Ref.~\cite{Taracchini2012}, their accumulated phase
  difference at merger is only $\sim0.5$ radians. Any NR phase
  error at merger larger than that, accumulated over $60$ cycles,
  would not improve the low-frequency accuracy of the EOB model.
  In fact, the $q=5$ and $q=8$ NR waveforms, despite being rather long and 
  accurate, do not provide new information to the EOB calibration.
  Truncation errors of these simulations dominate over other 
  numerical errors and EOB modeling errors. More accurate NR simulations of 
  large $q$ are therefore needed to further improve the low-frequency accuracy 
  of the EOB model. 

{\it Stability of the EOB calibration.} Although the differences
  among $\{\Acal{1}{N}(1),\Acal{2}{N}(1)\}$ waveforms can be
  distinguished by the global phase difference $\Delta\phi_g$, which is a
  highly sensitive quantity, it is not clear whether they can be
  distinguished by interferometric advanced detectors, such as LIGO. 
  Using the zero-detuned high-power advanced LIGO noise curve~\cite{Shoemaker2009} 
  and a total mass for the black-hole binary of $20M_\odot$, we quantify the data-analysis consequence
  of the differences between $\{\Acal{1}{N}(q),\Acal{1}{N}(q)\}$. Our
  study follows the procedure of Ref.~\cite{MacDonald:2012mp} and our
  results can be compared directly with those of
  Ref.~\cite{MacDonald:2012mp}.

First, we employ the quantity $||dh||/||h||$~\cite{Lindblom:2009ux} to measure the difference between waveforms 
$h_1$ and $h_2$, where $dh\equiv h_1-h_2$, $h=h_1$. The norm is defined through the inner-product
$\langle h_1,h_2\rangle\equiv 4 {\rm Re} \int_0^\infty(\tilde{h}_1(f)\tilde{h}^*_2(f))/{S_n(f)}\,df$
where $S_n(f)$ is the noise spectral density.

\begin{figure}[t]
  \includegraphics[width=0.37\textwidth]{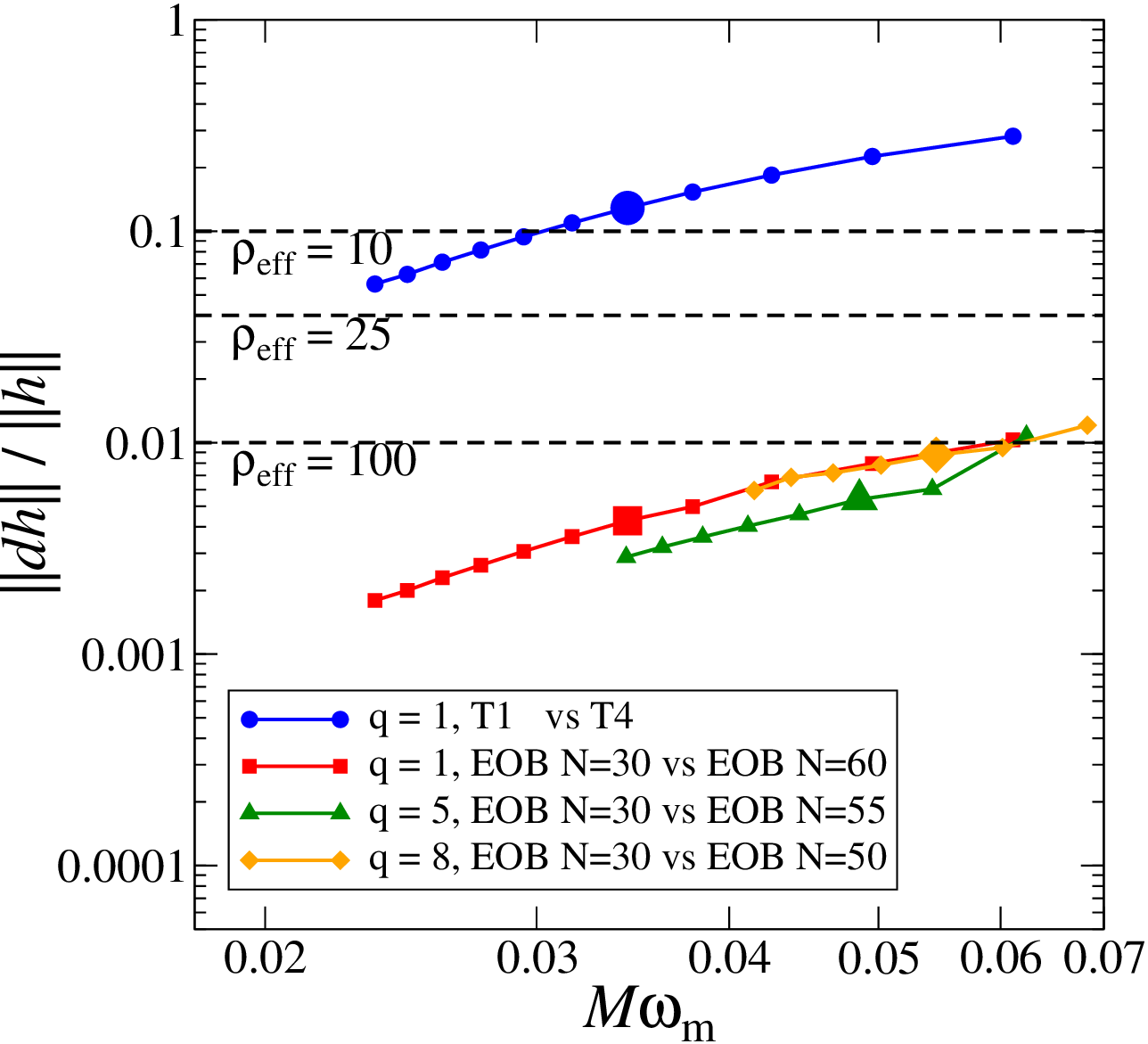}
  \caption{ \label{fig:dhh} We show $||dh||/||h||$ minimized over 
    time and phase of coalescence as a function of the hybrid matching
    frequency $\omega_m$ for EOB+NR hybrids where EOB waveforms
    are generated with the calibrated points
    $\{\Acal{1}{30}(q),\Acal{2}{30}(q)\}$ and $\{\Acal{1}{N_{\rm
        max}}(q),\Acal{2}{N_{\rm max}}(q)\}$. We also show the same quantity
    for PN+NR hybrids using TaylorT1 and TaylorT4 approximants. 
    The bigger symbol in each data set marks the matching
    frequency where the hybrid is built using $30$ cycles of NR waveforms.
    The horizontal lines mark the effective SNR $10$, $25$ and $100$, below
    which the difference between waveforms can not be distinguished by
    advanced LIGO detectors.
   }
\end{figure}
%
\begin{figure}
  \includegraphics[width=0.37\textwidth]{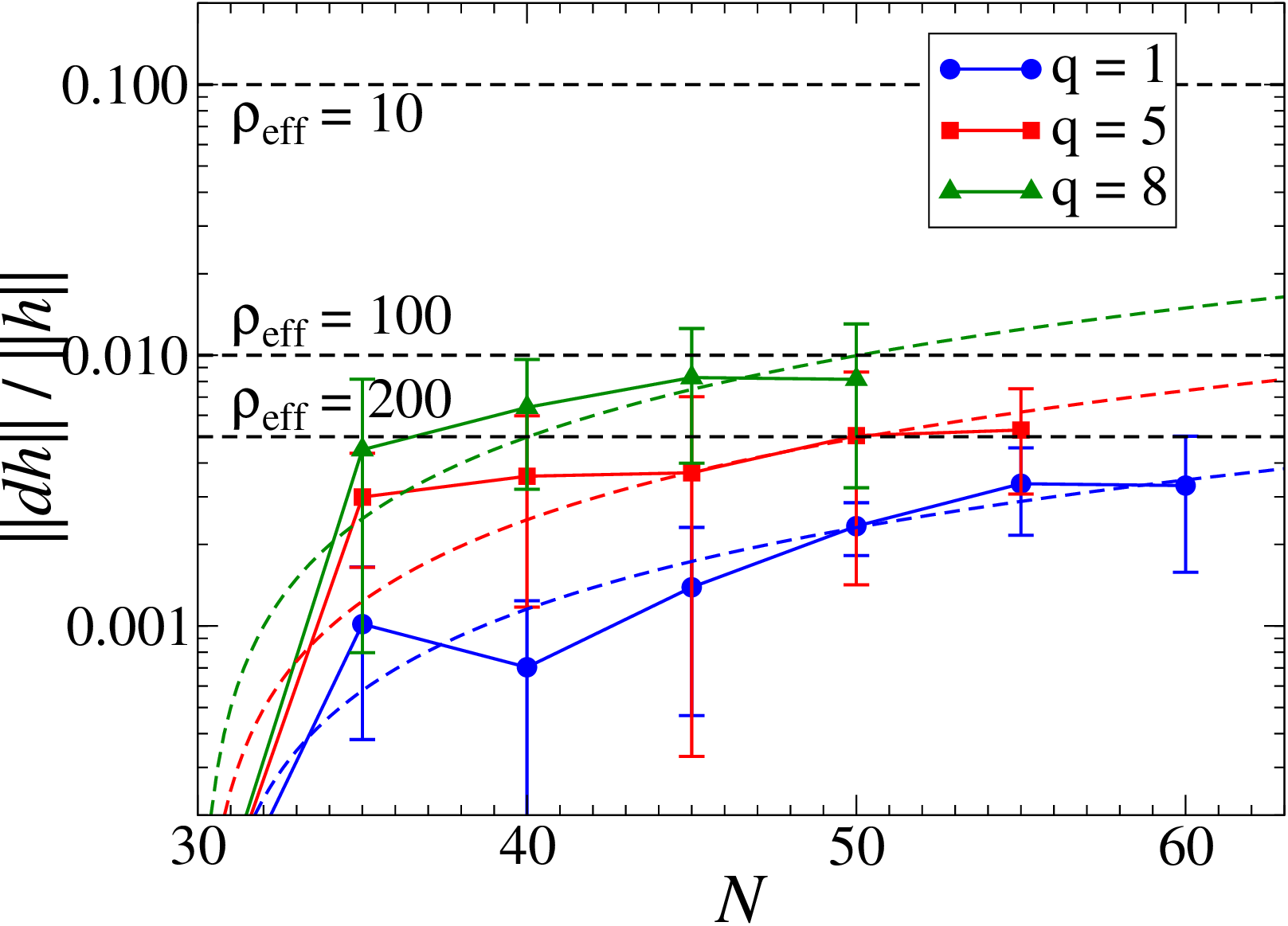}
  \caption{ \label{fig:dhh30} We show $||dh||/||h||$ between EOB
    waveforms generated with the calibrated points
    $\{\Acal{1}{30}(q),\Acal{2}{30}(q)\}$ and
    $\{\Acal{1}{N}(q),\Acal{2}{N}(q)\}$ as a function of the number of NR cycles $N$. We
    minimize $||dh||/||h||$ over time and phase of coalescence and
    use $30$ cycles of NR waveforms in the EOB+NR hybrids. The
    horizontal lines mark the effective SNR $100$ and $200$, below
    which the difference between waveforms can not be distinguished by
    advanced LIGO detectors.
   }
\end{figure}
When we minimize over time and phase of coalescence, as
well as physical parameters, $||dh||/||h||$  measures the relative loss of
signal-to-noise ratio (SNR). When we minimize over only the time and phase 
of coalescence, $||dh||/||h||$  
measures the bias in measuring source parameters, 
due to modeling errors. The bias is less than statistical errors when $||dh||/||h||<1/\rho_{\rm eff}$, where the effective SNR $\rho_{\rm  eff}=1/\epsilon\sqrt{n_D}\rho$ is proportional to the
single-detector SNR $\rho$ with a coefficient given by the number of
detectors $n_D$ and a safe factor $1/\epsilon$~\cite{Damour:2010} of
order unity. Satisfying this condition means that the detector cannot
distinguish $h_1$ and $h_2$. Either is an accurate enough template to 
measure the source parameters of the other. [We emphasize that the criterion of 
indistinguishability proposed in Ref.~\cite{Lindblom:2009ux}, i.e., $||dh||<1$, 
is a sufficient, but not necessary criterion, and it has been shown to be highly 
restrictive~\cite{Littenberg:2012uj}.]

In order to calculate $||dh||/||h||$, we need to complete the EOB
  inspiral waveforms $\{\Acal{1}{N_{\rm max}}(q),\Acal{2}{N_{\rm
      max}}(q)\}$ with merger and ringdown waveforms. Previous studies demonstrated 
that it is always possible to calibrate the EOB merger and ringdown
  waveforms to sufficient accuracy once the inspiral waveforms are accurately calibrated~\cite{PanEtAl:2011,Littenberg:2012uj}.  
So, here, we do not include the EOB merger and ringdown waveforms, but simply attach
  the NR late-inspiral, merger and ringdown waveforms to the EOB
  inspiral waveforms, starting at the matching frequency $\omega_m$, 
  i.e., we construct EOB+NR hybrid waveforms. This allows us to directly compare our results 
with the ones of  Ref.~\cite{MacDonald:2012mp}. In fact, for this reason, when building 
EOB + NR waveforms, we also follow the prescription of Ref.~\cite{MacDonald:2012mp} 
on the matching frequency, the time window for alignment and the choice of blending function.

In Fig.~\ref{fig:dhh}, we show $||dh||/||h||$ between
$\{\Acal{1}{30}(q),\Acal{2}{30}(q)\}$ and $\{\Acal{1}{N_{\rm
    max}}(q),\Acal{1}{N_{\rm max}}(q)\}$ waveforms as a function of the
matching frequency. We include also a comparison between PN+NR hybrid
waveforms constructed using the TaylorT1 and TaylorT4 approximants~\cite{Buonanno:2009} as
a validation of our code and to compare with Ref.~\cite{MacDonald:2012mp}. 
The difference between $\{\Acal{1}{30}(q),\Acal{2}{30}(q)\}$ and $\{\Acal{1}{N_{\rm
    max}}(q),\Acal{2}{N_{\rm max}}(q)\}$ EOB waveforms is more than an
order of magnitude smaller than the one obtained using the Taylor-PN approximants. 
Specifically, when attaching a $30$-cycle NR waveform at the end of the EOB inspiral waveform, the difference cannot be distinguished as long as  
$\rho_{\rm eff}<110$, which is an unlikely high
SNR for advanced detectors~\cite{MacDonald:2011ne}. This implies that nonspinning EOB
waveforms calibrated to $30$ or to $N_{\rm max}$ cycles of NR
waveforms are equivalent when searching for GWs and extracting binary parameters 
with advanced LIGO detectors. For the EOB model calibrated to $30$-cycle NR waveforms, we emphasize that the
implication of these results is not just the agreement of its waveform
with $N_{\rm max}$-cycle NR waveforms, but its agreement with the EOB
model calibrated to $60$-cycle NR simulations, i.e., the stability and
convergence of the calibrated EOB model up to $60$ cycles. Moreover, this 
result also demonstrates that calibrated higher-order PN terms (i.e., adjustable 
parameters) do not have a large effect at low frequency.

Can we extend this conclusion to $N>N_{\rm max}$? In
  Fig.~\ref{fig:dhh30}, we show $||dh||/||h||$ between EOB waveforms
  computed at the calibrated points $\{\Acal{1}{30}(q),\Acal{2}{30}(q)\}$ and
  $\{\Acal{1}{N}(q),\Acal{2}{N}(q)\}$ as function of $N$. We see that when $N$
  increases from $30$ to $N_{\rm max}$, $||dh||/||h||$ increases
  moderately from zero to $<1\%$ and the increase seems to be slowing down
  or becoming negative as we approach $N_{\rm max}$.  The oscillations in
  $||dh||/||h||$ are consistent with the NR error bars indicated in the plot and 
  estimated using the 12 different $\{\Acal{1}{N}(q),\Acal{2}{N}(q)\}$ points. 
  If we assume that the very mild increase of $||dh||/||h||$ is largely explained
  by NR errors, we might be tempted to conclude that the EOB model has converged beyond 
$N_{\rm max}$. However, we must be cautious in extrapolating the results. 
  Nevertheless, it is reasonable to expect that the variation of $||dh||/||h||$ per unit increase of $N$ eventually
  becomes a decreasing function of $N$ when $N$ is large enough, and
  consequently $||dh||/||h||$ becomes a concave function of $N$. We
  therefore obtain a conservative estimate of $||dh||/||h||$ by applying 
  a linear extrapolation of $||dh||/||h||$ that goes through $0$ at
  $N=30$ and best fit the data points. We find that $||dh||/||h||<0.05$
  until $N=370$, $235$ and $120$ for mass ratios $q=1$, $5$ and $8$,
  respectively. That is to say, when $\rho_{\rm eff}\le20$, EOB waveforms calibrated 
  to those numbers of NR cycles cannot be distinguished from
  EOB waveforms calibrated to $30$-cycle NR waveforms. 
One may hence generate $30$-cycle NR simulations to calibrate the EOB model, and use the calibrated model to produce EOB waveforms that are, for data analysis purposes, identical to NR waveforms of hundreds of cycles.

Finally, we compare these results to the length requirements of NR
waveforms set by previous works~\cite{Damour:2010,Boyle:2011dy,
  OhmeEtAl:2011, MacDonald:2011ne, MacDonald:2012mp} to guarantee the
accuracy of PN+NR hybrid waveforms for parameter
estimation. Basically, when NR simulations are sufficiently long,
their starting frequency $f_{\rm NR}$ can be reduced to $f_{\rm PN}$,
below which all PN waveform families and PN-based EOB model are
consistent. Direct estimates of the number of NR cycles before merger
required for accurate hybrid waveforms were made in
Ref.~\cite{Damour:2010} (see the table in Fig. 4 of
Ref.~\cite{Damour:2010}). When $\rho_{\rm eff}\le20$, for advanced
LIGO detectors, the number of GW cycles required for $q=1$, $4$ and
$10$ nonspinning NR simulations is $12$, $190$ and $1268$,
respectively. Combining those results with ours we conclude that when
$\rho_{\rm eff}\le20$ and $q\le5$ the nonspinning EOB waveforms of
{\it any} length are sufficiently accurate for parameter estimation
with advanced LIGO detectors. Note again that these EOB waveforms
are generated by the EOB model calibrated to {\it only} $30$-cycle NR
simulations.

{\it Conclusions.} We found that the EOB-model calibration  against NR
simulations is stable with respect to the length of NR simulations. In
the nonspinning limit with mass ratio $q\le 8$, the difference between
EOB waveforms calibrated against $30$-cycle and $\sim 60$-cycle NR
simulations can not be distinguished by advanced LIGO detectors 
when $\rho_{\rm eff}<110$. Extrapolating our results to a larger number of cycles, 
making rather conservative assumptions, which use the overstrict 
criterion from Ref.~\cite{Lindblom:2009ux}, we estimated that the
nonspinning EOB model calibrated to existing NR simulations is
sufficiently accurate for advanced-LIGO parameter estimation when
$\rho_{\rm eff}<20$, $q<5$ and $M \geq 20 M_\odot$. Moreover, since 
EOB waveforms overcome the frequency gap, they can completely 
replace PN + NR hybrid waveforms~\cite{Boyle:2011dy,OhmeEtAl:2011, MacDonald:2011ne, 
MacDonald:2012mp}. Extending this conclusion to larger
$\rho_{\rm eff}$ or $q$ requires longer and more accurate NR
simulations. We plan in the near future to extend this kind of study to 
the spinning EOB model~\cite{Taracchini2013}.
We expect that in the presence of spins, we might need 
longer and more accurate NR simulations, especially 
in the extremal-spin limit, but the length can be much less than those suggested    
by previous studies that aimed at reducing $f_{\rm NR}$ to $f_{\rm PN}$. 

{\it Acknowledgments}
  A.B., Y.P. and A.T. acknowledge partial support from NSF Grants
  No. PHY-0903631 and No. PHY-1208881, and NASA Grant
  NNX09AI81G. A.T.. acknowledges support also from the Maryland Center for
  Fundamental Physics.  A.M. and H.P acknowledge support from NSERC of
  Canada, from the Canada Research Chairs Program, and from the
  Canadian Institute for Advanced Research.
  M.S. and B.S. gratefully acknowledge support from the Sherman Fairchild
  Foundation and from NSF grants PHY-106881 and PHY-1005655 at Caltech.
  M.B. and L.K.  gratefully acknowledge support from the Sherman Fairchild
  Foundation and from NSF grants PHY-1306125 and PHY-1005426 at Cornell.
  The numerical relativity simulations were performed at the GPC
  supercomputer at the SciNet HPC Consortium; SciNet is
  funded by: the Canada Foundation for Innovation (CFI) under the
  auspices of Compute Canada; the Government of Ontario; Ontario
  Research Fund--Research Excellence; and the University of
  Toronto. Further computations were performed on the Caltech compute
  cluster Zwicky, which was funded by the Sherman Fairchild Foundation
  and the NSF MRI-R2 grant No. PHY-0960291.

\vspace{-1cm}

%
\end{document}